\documentclass[twocolumn,showpacs,amscd,amsmath,amssymb,verbatim]{revtex4}

\usepackage{graphicx}

\usepackage{textcomp}
\usepackage[OT1,T1]{fontenc}

\begin{document}

\title{Influence of magnetic interaction between impurity and impurity-liberated
spins on the magnetism in the doped Haldane chain compounds PbNi${}_{2-\mathit{x}}${\itshape
A}${}_{\mathit{x}}$V${}_{2}$O${}_{8}$ ({\itshape A} = Mg, Co)}
\author{Andrej Zorko}
\email{andrej.zorko@ijs.si}
\affiliation{Jo\v{z}ef Stefan Institute, Jamova 39, 1000 Ljubljana,
Slovenia}\author{Denis Ar\v{c}on}
\affiliation{Jo\v{z}ef Stefan Institute, Jamova 39, 1000 Ljubljana,
Slovenia}\author{Alexandros Lappas}
\affiliation{Institute of Electronic Structure and Laser, Foundation
for Research and Technology -- Hellas, P.O. Box 1527, 71110 Heraklion,
Crete, Greece}\author{Zvonko Jagli\v{c}i\'c}
\affiliation{Institute of Mathematics, Physics and Mechanics, Jadranska
19, 1000 Ljubljana, Slovenia}\label{I1}\label{I2}\label{I3}
\date{\today}
\begin{abstract}
A comprehensive study of impurity-induced magnetism in nonmagnetically
(Mg${}^{2+}$) and magnetically (Co${}^{2+}$) doped PbNi${}_{2}$V${}_{2}$O${}_{8}$
compounds is given, using both macroscopic dc susceptibility and
local-probe electron spin resonance (ESR) techniques. Magnetic coupling
between impurity-liberated spins is estimated from a linewidth of
low-temperature ESR signal in Mg-doped samples. In addition, in
the case of magnetic cobalt dopants the impurity-host magnetic exchange
is evaluated from the Co-induced contribution to the linewidth in
the paramagnetic phase. The experimentally observed severe broadening
of the ESR lines in the magnetically doped compounds with respect
to nonmagnetic doping is attributed to a rapid spin-lattice relaxation
of the Co${}^{2+}$ ions, which results in a bottleneck-type of temperature
dependence of the induced linewidth. The exchange parameters obtained
from the ESR analysis offer a satisfactory explanation of the observed
low-temperature magnetization in doped samples. 
\end{abstract}
\maketitle

\section{Introduction}

Haldane integer-spin chains with antiferromagnetic coupling have
been extensively studied experimentally as well as theoretically
in the last two decades. This is due to their fascinating property
conjectured by Haldane \cite{HaldanePRL50}. Namely, in contrast
to half-integer-spin chains, in integer-spin chains a quantum disordered
singlet ground state with correlations decaying exponentially is
separated from the lowest excited state (spin gap). Such character
of the Haldane chains with only the nearest-neighbor ({\itshape
nn}) isotropic antiferromagnetic exchange coupling was satisfactory
accounted for by the valence-bond-solid model \cite{AffleckPRL59},
which introduced valence bonds emerging and terminating at neighboring
sites. The validity of this model was experimentally confirmed for
the first time in the spin $S=1$ compound NENP by observing $S=1/2$
liberated end-chain spins when a portion of the valence bonds was
intentionally broken by introducing impurities to partially replace
the $S=1$ spins \cite{HagiwaraPRL65}.

In general, impurities have been in the past often deliberately
introduced to host materials and thus employed to reveal the magnetic
character of the host. The instructive examples cover a variety
of different low-dimensional quantum spin systems including high-{\itshape
T}${}_{\mathit{c}}$ superconductors \cite{HudsonNature411}. Moreover,
the impurities can dramatically affect the ground state of the host
material leading to unexpected magnetic phenomena. One of their
most astonishing consequences is the induction of long-range magnetic
ordering upon doping, an effect known as ``order-by-disorder effect''
\cite{VillainJP41}, where disorder in a form of random doping causes
magnetic ordering in the host material. This mechanism is known
to set-in in different spin-gap systems including the spin-Peierls
compound CuGeO${}_{3}$ substitutionally doped with different nonmagnetic
\cite{HasePRL71,LussierJPCM7,MasudaPRL80} or magnetic ions \cite{LussierJPCM7,AndersonPRB56},
the vacancy doped two-leg spin-ladder compound SrCuO${}_{3}$ \cite{AzumaPRB55}
and the three-dimensional coupled-spin-dimer system TlCuCl${}_{3}$
doped with nonmagnetic impurities \cite{OOsawaPRB66}. Recently,
also the first Haldane-chain compound PbNi${}_{2}$V${}_{2}$O${}_{8}$
that undergoes a transition to a magnetically ordered ground state
at low temperatures when doped with either nonmagnetic \cite{UchiyamaPRL83} 
Mg${}^{2+}$ or magnetic \cite{UchinokuraJMMM226,ImaiCM2004} Co${}^{2+}$, 
Cu${}^{2+}$ and Mn${}^{2+}$ ions on Ni${}^{2+}$
($S=1$) sites, was reported. There are several common features of
the phase transitions in the above-mentioned systems, the first
one being that already a very small amount of impurities induces
long-range ordering. Second, it appears universal that the phase-transition
temperature increases as a function of the doping level at low concentrations
and decreases at higher concentrations \cite{MasudaPRL80,AzumaPRB55,OOsawaPRB66,ImaiCM2004,UchinokuraPhysicaB}.
Therefore, a quest to find a unified pictures, which would satisfactory
explain the impurity-induced long-range ordering in spin-gap systems,
is currently underway.

The way to magnetic ordering in the spin-gap systems is paved by
clusters of exponentially decaying staggered moments induced next
to the impurity sites \cite{LaukampPRB57}, which are magnetically
coupled through the gaped medium. Weak coupling results in in-gap
energy states, which dominate the low-temperature magnetic character
of the doped systems. However, the picture of the impurity-induced
magnetic ordering in spin-gap systems still remains rather unclear.
In particular, the mechanisms leading to the development of three-dimensional
intercluster spin correlations are particularly elusive. In order
to elucidate this intriguing phenomenon it is crucial to develop
a deeper understanding of the magnetic interactions within the pockets
of the impurity-induced staggered moments as well as between them.

Therefore, we decided to act upon a comprehensive magnetic investigation
of the Haldane chain compound PbNi${}_{2}$V${}_{2}$O${}_{8}$ doped
with nonmagnetic Mg${}^{2+}$ and magnetic Co${}^{2+}$ impurities.
The study incorporates macroscopic dc susceptibility measurements
and local-scale magnetic resonance techniques, which already proved
to yield invaluable information on the impurity-induced magnetism
in Mg-doped compounds \cite{ZorkoPRB65,ArconEPL65}. The results
of the electron spin resonance (ESR) are displayed in the present
paper while the findings of the complementary nuclear magnetic resonance
(NMR) measurements on {}\textsuperscript{51}V nuclei weakly coupled
to electronic moments, are presented elsewhere \cite{ZorkoPreprint}.
Combining the results of the dc susceptibility measurements and
the ESR technique, we are able to evaluate the magnetic coupling
within the impurity-induced pockets of staggered moments as well
as to highlight the role of the intercluster coupling on the long-range
magnetic ordering.

\section{EXPERIMENTAL DETAILS}

Polycrystalline samples were prepared by a solid-state reaction
with the details of the procedure already been published \cite{LappasPRB66}.
The efficiency of the Mg${}^{2+}$ and Co${}^{2+}$ cations replacements
for Ni${}^{2+}$ ions was experimentally verified by powder x-ray
and neutron diffraction \cite{MastorakiAPA74,MastorakiJSSC177}.
The presence of any impurity phases was limited below the sensitivity
level of the x-ray diffractometer. dc magnetization measurements
were performed on a Quantum Design SQUID magnetometer in a magnetic field
of 100~mT down to 2~K. For ESR measurement a Bruker E580 FT/CW spectrometer
was used. These measurements were conducted between room temperature
and 5~K at a Larmor frequency of $\nu _{L}\approx 9.4$ GHz (X-band).

\section{EXPERIMENTAL RESULTS}

\subsection{dc susceptibility}
\begin{figure}[ht]
\begin{center}
\includegraphics[angle=0,width=7.2cm,scale=1.0]{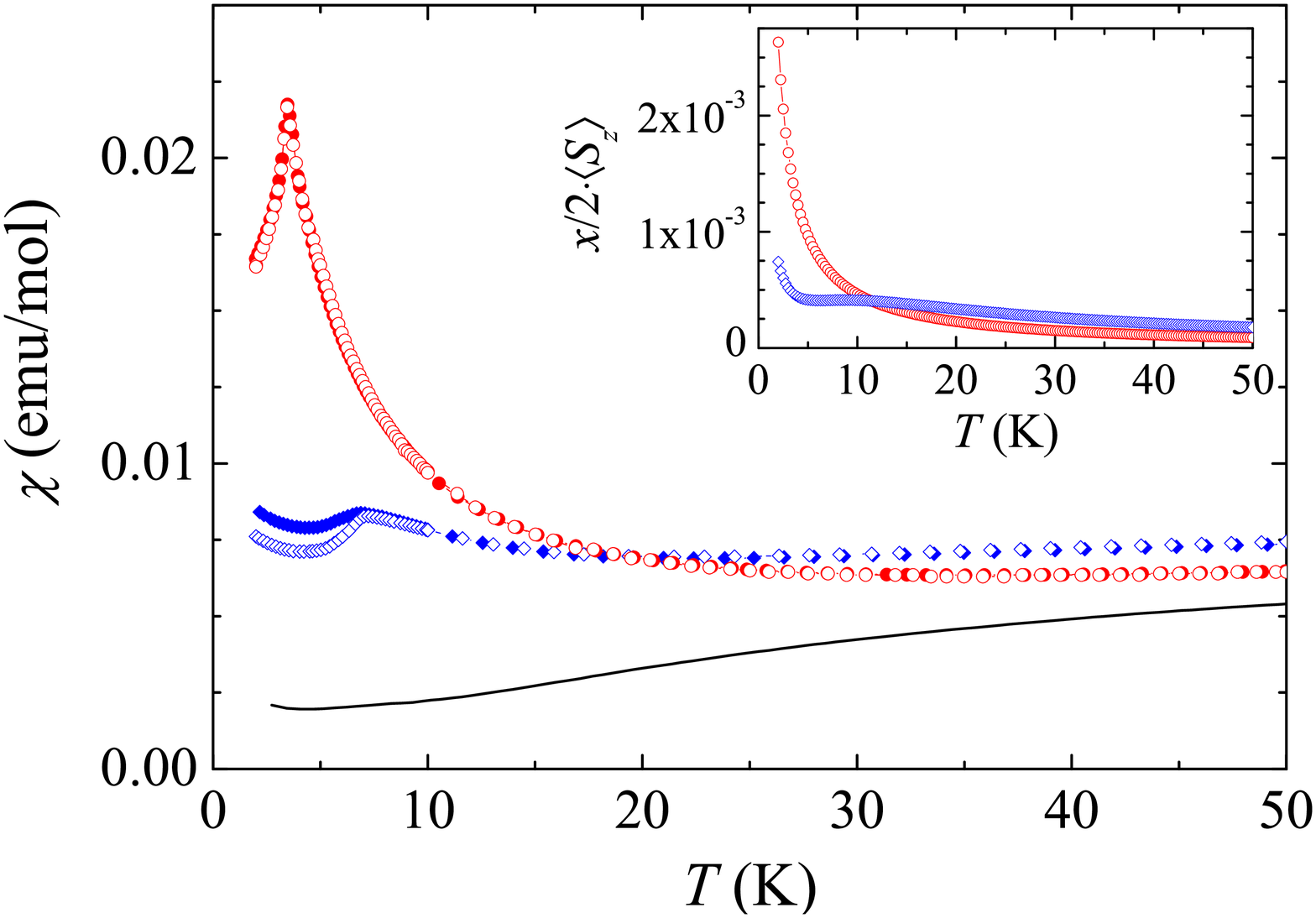}

\end{center}
\caption{(Color online) dc magnetic susceptibility of the PbNi${}_{1.88}$Mg${}_{0.12}$V${}_{2}$O${}_{8}$
(circles) and PbNi${}_{1.92}$Co${}_{0.08}$V${}_{2}$O${}_{8}$ (diamonds)
compounds in the field cooling (solid symbols) and zero-field cooling
(open symbols) regime in the magnetic field of 100~mT. The solid
line is displaying the spin-gap behavior of the pristine PbNi${}_{2}$V${}_{2}$O${}_{8}$
material. Inset shows a temperature evolution of the average impurity-induced
spin size for $S=0$ (circles) and $S=3/2$ (diamonds) impurities
in a simplified model described in subsection \ref{XRef-Section-52103829}.}
\label{XRef-Figure-42710958}
\end{figure}

As previously reported, doping the PbNi${}_{2}$V${}_{2}$O${}_{8}$
compound with either nonmagnetic Mg${}^{2+}$ or magnetic Co${}^{2+}$
impurities results in the long-range magnetic ordering at low temperatures
\cite{UchiyamaPRL83,UchinokuraPhysicaB}. Evidence to the nature
of the ground state arise from the presence of sharp magnetic Bragg
peaks in the neutron powder patterns \cite{LappasPRB66,MastorakiJSSC177}.
The onset of the magnetic ordering is clearly expressed in the characteristic
peaks of the magnetic susceptibility curves at low temperatures
as shown in Fig. \ref{XRef-Figure-42710958}. The transition temperatures
in the external magnetic field of 100~mT are around 3.4~K and 7.1~K 
in the case of PbNi${}_{1.88}$Mg${}_{0.12}$V${}_{2}$O${}_{8}$
and PbNi${}_{1.92}$Co${}_{0.08}$V${}_{2}$O${}_{8}$, respectively.

The first rather unusual feature of the observed phase transition
is the significantly different value of the phase-transition temperature
in both materials, although the stoichiometric amounts of impurities
are similar. Such an increase of the transition temperature in the
case of cobalt doping compared to nonmagnetic doping was not observed
in the immensely studied CuGeO${}_{3}$ compound, which is a prototypical
system for impurity-induced magnetic ordering \cite{HasePRL71,LussierJPCM7,MasudaPRL80,AndersonPRB56}.
However, it is in line with the observation that the magnetically
ordered state is much more stable against the external magnetic
field in Co-doped PbNi${}_{2}$V${}_{2}$O${}_{8}$ than in Mg-doped
samples, as experimentally verified by dc magnetization and NMR
measurement in magnetic fields of several Tesla \cite{ZorkoPreprint}
\begin{figure}[ht]
\begin{center}
\includegraphics[angle=0,width=7.2cm,scale=1.0]{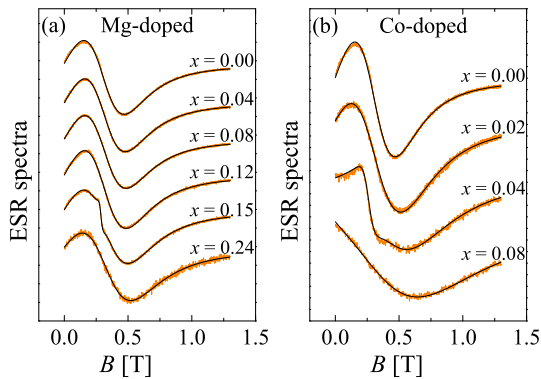}

\end{center}
\caption{(Color online) The evolution of the room-temperature X-band
ESR spectra of PbNi${}_{2-\mathit{x}}${\itshape A}${}_{\mathit{x}}$V${}_{2}$O${}_{8}$
with increased level of doping in the case of (a) nonmagnetic $A={\mathrm{Mg}}^{2+}$
and (b) magnetic $A={\mathrm{Co}}^{2+}$ dopants. The solid lines
represent fits to broad Lorentzian distribution as explained in
the text.}
\label{XRef-Figure-427115449}
\end{figure}

Apart from the significant increase of the transition temperature
in the Co-doped compound, another feature is noticeable in Fig.
\ref{XRef-Figure-42710958}. Namely, while the Mg-doped sample exhibits
no hysteresis between the zero-field cooling (ZFC) and field cooling
(FC) regimes in the magnetically ordered state, such behavior is
observed in the case of the Co-doped sample. This experimental finding
seems unusual in the light of the neutron diffraction performed
on both compounds in the magnetically ordered state, suggesting
a magnetic ordering of similar character. More specifically, purely
magnetic neutron diffraction patterns indicate that the average
magnetic moment per Ni site pointing along the direction of the
Ni chains ({\itshape c} axis) is in both compounds close to $1\mu
_{B}$ \cite{MastorakiAPA74,MastorakiJSSC177}. These measurements,
however, can not yield information on the components of the magnetic
moments perpendicular to the chains due to the uniaxial symmetry
of the lattice and the powder nature of the samples. It has been
recently argued by Imai et al. \cite{ImaiCM2004} that the observed
irreversibility between the ZFC and FC measurements is not a signature
of a spin-glass behavior but should rather be due to the occurrence
of weak ferromagnetism in Co-doped samples, attributed to the presence
of the Dzyaloshinsky-Moriya (DM) interaction \cite{DzyaloshinskyJPCS4,MoriyaPR120}.
Such antisymmetric anisotropic interaction is indeed present between
Ni spins \cite{ZorkoThesis}, however, the general anisotropic exchange
between impurity and host spins should be involved in the ZFC/FC
irreversibility.

Although the nominal concentrations of impurities in both samples
are similar, the low-temperature peak in the magnetic susceptibility
is much more pronounced in the Mg-doped sample despite the fact
that Mg dopants are nonmagnetic. The uniform static spin susceptibility
can be obtained from the dynamical spin structure factor $S^{zz}(
\textbf{q},\omega ) =\mathrm{Re}\int _{0}^{\infty } e ^{i\omega
t}\langle S_{\textbf{q}}^{z}( t) S_{-\textbf{q}}^{z}( 0) \rangle
dt$, reflecting the distribution of the spectral weight of spin
excitations (the brackets $\langle ...\rangle $ denote statistical
averaging). Kramers-Kronig relations give the following expression
\begin{equation}
\chi _{0}\propto \frac{1}{\pi }\operatorname*{\lim }\limits_{\textbf{q}\rightarrow
\:0}\mathcal{P}\operatorname*{\int }\limits_{-\infty }^{\infty }\left(
1- e ^{-\hbar \omega /k_{B}T}\right) \frac{S^{zz}( \textbf{q},\omega
) }{\omega }d\omega ,
\end{equation}
\begin{figure}[ht]
\begin{center}
\includegraphics[angle=0,width=7.2cm,scale=1.0]{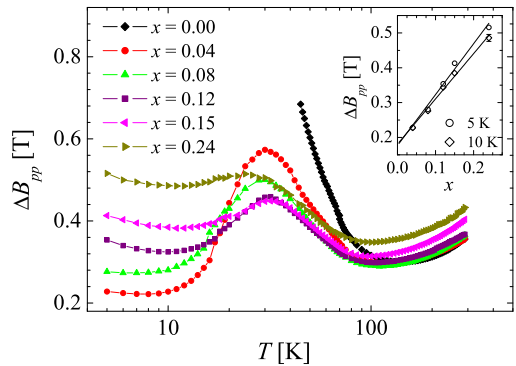}

\end{center}
\caption{(Color online) The temperature dependence of the X-band
ESR linewidth in PbNi${}_{2-\mathit{x}}$Mg${}_{\mathit{x}}$V${}_{2}$O${}_{8}$
compounds. Inset shown a linear dependence of the impurity-induced
line broadening at 5~K and 10~K.}
\label{XRef-Figure-427122636}
\end{figure}

\noindent where $ \mathcal{P}$ stands for the Cauchy principle value.
Bering in mind that the static susceptibility is dominated by low-energy
spin excitations, the intensities of the low-temperature magnetic
susceptibility give a clear signal that the in-gap impurity-induced
density of states is peaked at much lower frequencies when magnesium
impurities are present. This is not surprising, as stronger magnetic
coupling is expected in the case of cobalt impurities, which shifts
the in-gap states to higher energies. Moreover, our analysis based
on the parameters obtained from the ESR results, gives also a quantitative
agreement, as demonstrated in the subsection \ref{XRef-Section-52103829}.

\subsection{Electron spin resonance}

As previously reported, X-band ESR absorption spectra of the PbNi${}_{2}$V${}_{2}$O${}_{8}$
are fairly broad at room temperature and further broaden with lowering
the temperature \cite{ZorkoPRB65}. Partial substitution of the magnetic
Ni${}^{2+}$ ions with nonmagnetic Mg${}^{2+}$ ions results in additional
broadening of the resonance lines, which moderately increases with
the level of doping as clearly observed in Fig. \ref{XRef-Figure-427115449}a.
On the other hand, the case of Co-doping results in extreme broadening
of the ESR spectra (Fig. \ref{XRef-Figure-427115449}b), reflecting
the magnetic nature of the dopants.

All the spectra can be satisfactory fitted with a broad Lorentzian
function taking into account the resonant absorption at positive
as well as negative resonant field. In addition, some samples require
an extra narrow component, whose intensity proves to be sample dependent
even within the same nominal stoichiometry. This fact allows us
to attribute this additional signal to impurity phases present in
the sample. However, judged from very small relative intensities
of the spurious signals (i.e., below 1\%) such impurities would
be unobservable in x-ray diffraction patterns. 
\begin{figure}[ht]
\begin{center}
\includegraphics[angle=0,width=7.2cm,scale=1.0]{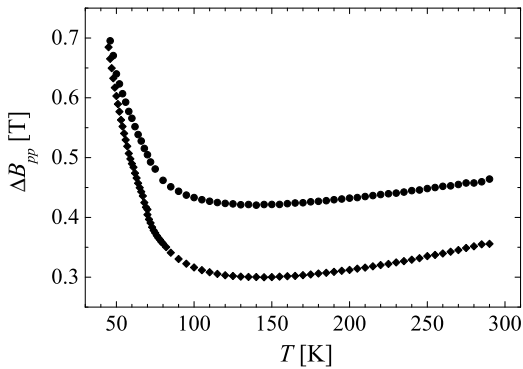}

\end{center}
\caption{The temperature dependence of the ESR linewidth in PbNi${}_{1.98}$Co${}_{0.02}$V${}_{2}$O${}_{8}$
(circles) compared to PbNi${}_{2}$V${}_{2}$O${}_{8}$ (diamonds).}
\label{XRef-Figure-427125441}
\end{figure}

With lowering temperature a diverse nature of the ESR spectra in
Mg-doped samples becomes evident (see Fig. \ref{XRef-Figure-427122636}).
Based on the scaling of the low-temperature ESR intensities with
the doping level and the temperature evolution of the observed {\itshape
g}-factor shifts \cite{ZorkoPRB65}, we were previously able to attribute
the additional low-temperature signal in vacancy-doped samples to
impurity-liberated end-chain degrees of freedom, which are delocalized
on the scale of the correlation length $\xi$ \cite{KimEPJB4}. It
was further argued that these end-chain spins at both sides of a
particular impurity were ferromagnetically coupled through interchain
exchange providing a mechanism for three-dimensional magnetic ordering.
Such ferromagnetic coupling was indeed later experimentally confirmed
by specific heat measurements reported by Masuda et al. \cite{MasudaPRB66}.
The exact origin of the rather broad low-temperature ESR signals
still, however, remains to be revealed. The ESR linewidth reflects
the strength of the magnetic coupling between liberated spins, which
makes its analysis invaluable for the understanding of the impurity-induced
magnetism.

In Co-doped samples we were able to follow the temperature evolution
of the spectra below room temperature only in the case of the sample
with the lowest doping level, i.e., in PbNi${}_{1.98}$Co${}_{0.02}$V${}_{2}$O${}_{8}$.
The ESR linewidth of this compound is compared to the linewidth
of the pristine compound in Fig. \ref{XRef-Figure-427125441}. It
exhibits a rather different behavior from the case of the Mg-doped
materials as the linewidths of the two samples are clearly converging
towards the same values when decreasing the temperature. To explain
the extreme broadening of the ESR spectra due to Co${}^{2+}$ ions
and a bottleneck-type of the impurity-induced contribution to the
ESR linewidth, the spin nature of the dopants is successfully incorporated
into the analysis in the next section.

\section{ANALYSIS AND DISCUSSION}

\subsection{ESR measurements in the paramagnetic phase }

To rationalize the observed broadening of the ESR spectra at room
temperature due to doping (Fig. \ref{XRef-Figure-427115449}), we
presume that the impurities do not have any noticeable effect on
the crystal structure of the materials \cite{MastorakiAPA74,MastorakiJSSC177}.
Since the dominant spin anisotropy contributions in the investigated
materials are of single-ion anisotropy type \cite{ZheludovPRB62}
and DM type \cite{ZorkoThesis}, it can be assumed that they do not
change appreciably at Ni sites when dopants are introduced. Therefore,
the impurities have to influence the time-evolution of spin correlation
functions entering the expression of the ESR linewidth. The peak-to-peak
linewidth of the Lorentzian line in the exchange-narrowing limit
is given by \cite{CastnerPRB4}
\begin{equation}
\Delta  B_{pp}={\left. \frac{C}{g \mu _{B}}\left( \frac{{M_{2}}^{3}}{M_{4}}\right.
\right) }^{1/2}.
\end{equation}

\noindent Here {\itshape C} denotes a constant of the order of unity
and {\itshape g} the {\itshape g}-factor, while the second and the
forth moment of the absorption lines are given as
\begin{align}
\label{XRef-Equation-430231550}%
M_{2}&=\frac{\left\langle  \left[ \mathcal{H}^{\prime },M^{+}\right]
[ M^{-},\mathcal{H}^{\prime }] \right\rangle  }{\left\langle  M^{+}M^{-}\right\rangle
},
\\%
\label{XRef-Equation-43023168}%
M_{4}&=\frac{\left\langle  \left[ \mathcal{H}-\mathcal{H}_{Z},\left[
\mathcal{H}^{\prime },M^{+}\right] \right] [ \mathcal{H}-\mathcal{H}_{Z},\left[
\mathcal{H}^{\prime },M^{-}\right] ] \right\rangle  }{\left\langle
M^{+}M^{-}\right\rangle  }.
\end{align}

\noindent The Hamiltonian $\mathcal{H}=\mathcal{H}_{0}+\mathcal{H}^{\prime
}$ of the system is conventionally divided into two parts. The main
part $\mathcal{H}_{0}=\mathcal{H}_{Z}+\mathcal{H}_{e}$ contains
only the Zeeman Hamiltonian $\mathcal{H}_{Z}$ and the exchange Hamiltonian
$\mathcal{H}_{e}$ of the host. The perturbative part $\mathcal{H}^{\prime
}$ includes all the anisotropy contributions of the host as well
as the impurity Hamiltonian $\mathcal{H}^{i}=\mathcal{H}_{Z}^{i}+\mathcal{H}_{e}^{i-h}+\mathcal{H}_{LS}$
\cite{NagataJPSJ44}, where $\mathcal{H}_{Z}^{i}$ denotes the Zeeman
Hamiltonian of the impurity system, $\mathcal{H}_{ex}^{i-h}$ the
impurity-host exchange coupling and $\mathcal{H}_{LS}$ the impurity
spin-orbit coupling, all these terms being nonzero in the case of
doping with magnetic dopants.

\subsubsection{Mg-doping}

\begin{figure}[ht]
\begin{center}
\includegraphics[angle=0,width=7.2cm,scale=1.0]{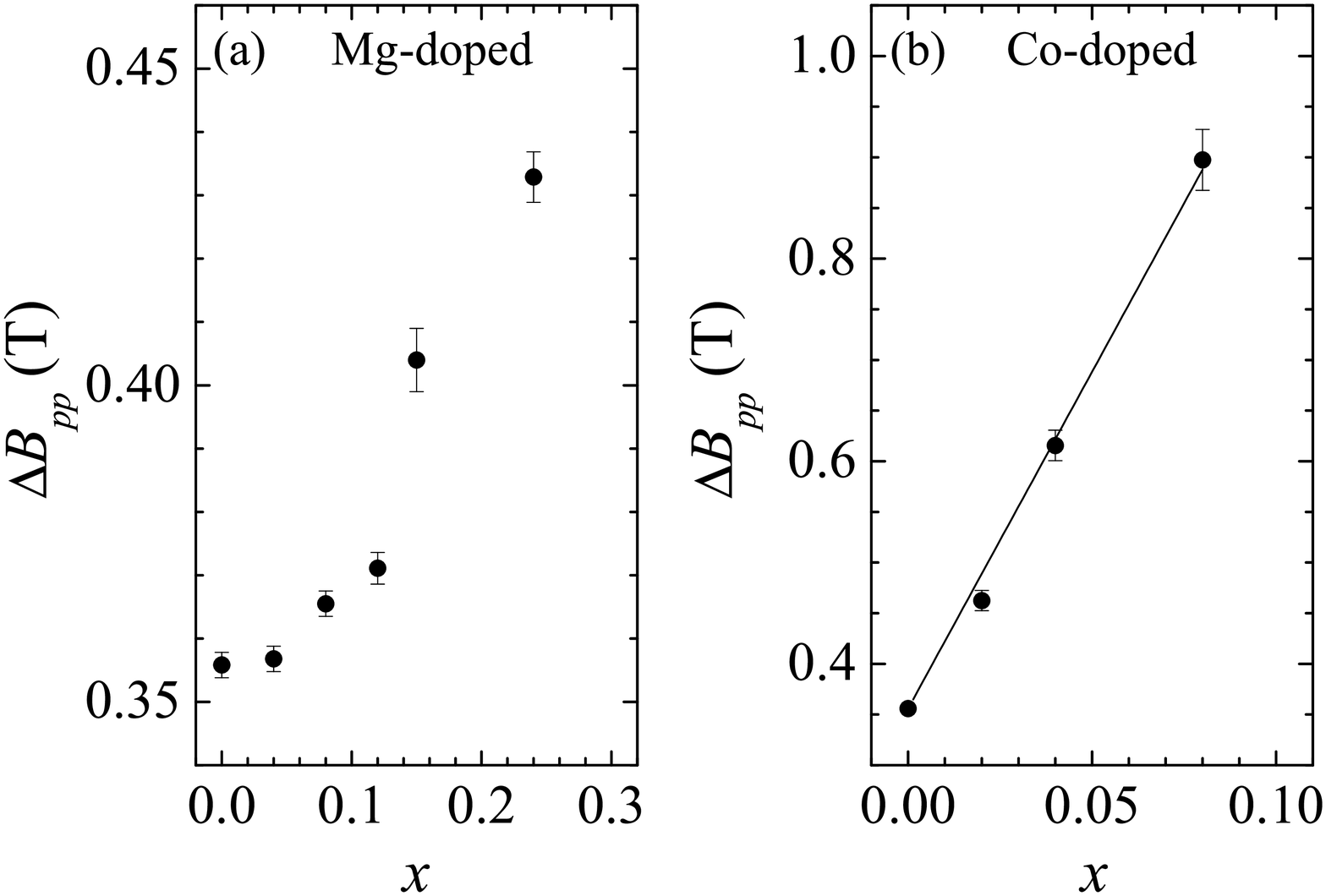}

\end{center}
\caption{Linewidth of the X-band ESR spectra recorded at room temperature
in (a) Mg-doped and (b) Co-doped PbNi${}_{2}$V${}_{2}$O${}_{8}$.}
\label{XRef-Figure-4271361}
\end{figure}

In fact, the spin correlations can be significantly affected by
impurities \cite{HonePRB72} in low-dimensional systems, where the
exchange pathways are severely limited and the diffusional decay
of spin correlations at longer times becomes important \cite{RichardsEnricoFermi}.
The rate of spin diffusion across the impurity site depends on the
size of the impurity spin and the impurity-host exchange coupling
\cite{RichardsPRB10}. Nonmagnetic impurities act as reflecting agents,
disabling the spin polarization to diffuse away and thus effectively
diminishing the exchange-narrowing mechanism. A significant impurity-induced
broadening of the ESR absorption lines is thus expected \cite{RichardsEnricoFermi}.
However, the observed broadening, presented in Fig. \ref{XRef-Figure-4271361}a,
is not that prominent. This experimental finding can be explained
by the relatively large interchain exchange coupling $J_{1}$ in
the case of the PbNi${}_{2}$V${}_{2}$O${}_{8}$ compound, $|{zJ}_{1}/J|\approx
$0.03, where $J=95$~K (in units of {\itshape k}{\itshape ${}_{B}$})
represents the dominant {\itshape nn} intrachain exchange \cite{ZheludovPRB62}.
Hence the spin system behaves as three-dimensional on the X-band
ESR time-scale since the out-of-chain diffusion rate, given by $J_{1}/\hbar
$, is larger than the ESR frequency. The inhibited diffusion along
the chains thus only partially influences the decay of spin correlations,
which is also in line with the Lorentzian lineshape of the recorded
ESR spectra \cite{RichardsEnricoFermi}.

\subsubsection{Co-doping}

Significantly more pronounced is the broadening of the ESR spectra
in the case of doping the parent PbNi${}_{2}$V${}_{2}$O${}_{8}$
compound with magnetic Co${}^{+2}$ ions (see Fig. \ref{XRef-Figure-4271361}).
We attribute the observed ESR signals solely to magnetic moments
of the host (Ni${}^{2+}$) spin system. Below-presented analysis shows that
Co${}^{2+}$ magnetic moments are strongly affected by the spin-lattice
relaxation at the room temperature, which severely broadens the
ESR spectra.

When the Co${}^{2+}$ ion (3{\itshape d}{}\textsuperscript{7} configuration)
is placed into a crystal field, which arises from octahedrally coordinated
O${}^{2-}$ ions, the ground orbital state is a triplet. Taking into
account the spin $S=3/2$ of the ``high-spin'' cobalt results in a twelve-fold
degenerated ground state. The spin-orbit coupling and a distortion
away from the cubic symmetry, as is the case in PbNi${}_{2}$V${}_{2}$O${}_{8}$
\cite{MastorakiJSSC177}, splits the energy levels, however, there
remains a significant amount of orbital moment in the ground state
\cite{PilbrowCo2+}. The spin dynamics of Co${}^{2+}$ ions are strongly
affected by lattice vibrations through the spin-orbit coupling.
Extreme broadening of ESR lines due to a presence of small concentrations
of Co${}^{2+}$ ions was reported in the past in several systems
\cite{NagataJPSJ44,GulleyPRB6,vanderVlistPRB30}. As Co${}^{2+}$
is a strongly relaxing ion, the broadening can be attributed to
an interplay between impurity-host cross relaxation and spin-lattice
relaxation of the impurity. The effect of the impurities is determined
by the relative rate of the magnetic energy flow from the host to
the impurity system with respect to the rate of the energy flow
between the impurities and the underlying lattice. The expected
ESR linewidth can then be expressed by the equation \cite{GulleyPRB6}
\begin{equation}
\Delta  B_{pp}=\Delta  B_{0}+\frac{\eta }{1+\eta }\Delta  B_{i},%
\label{XRef-Equation-43020556}
\end{equation}
\begin{figure}[ht]
\begin{center}
\includegraphics[angle=0,width=7.2cm,scale=1.0]{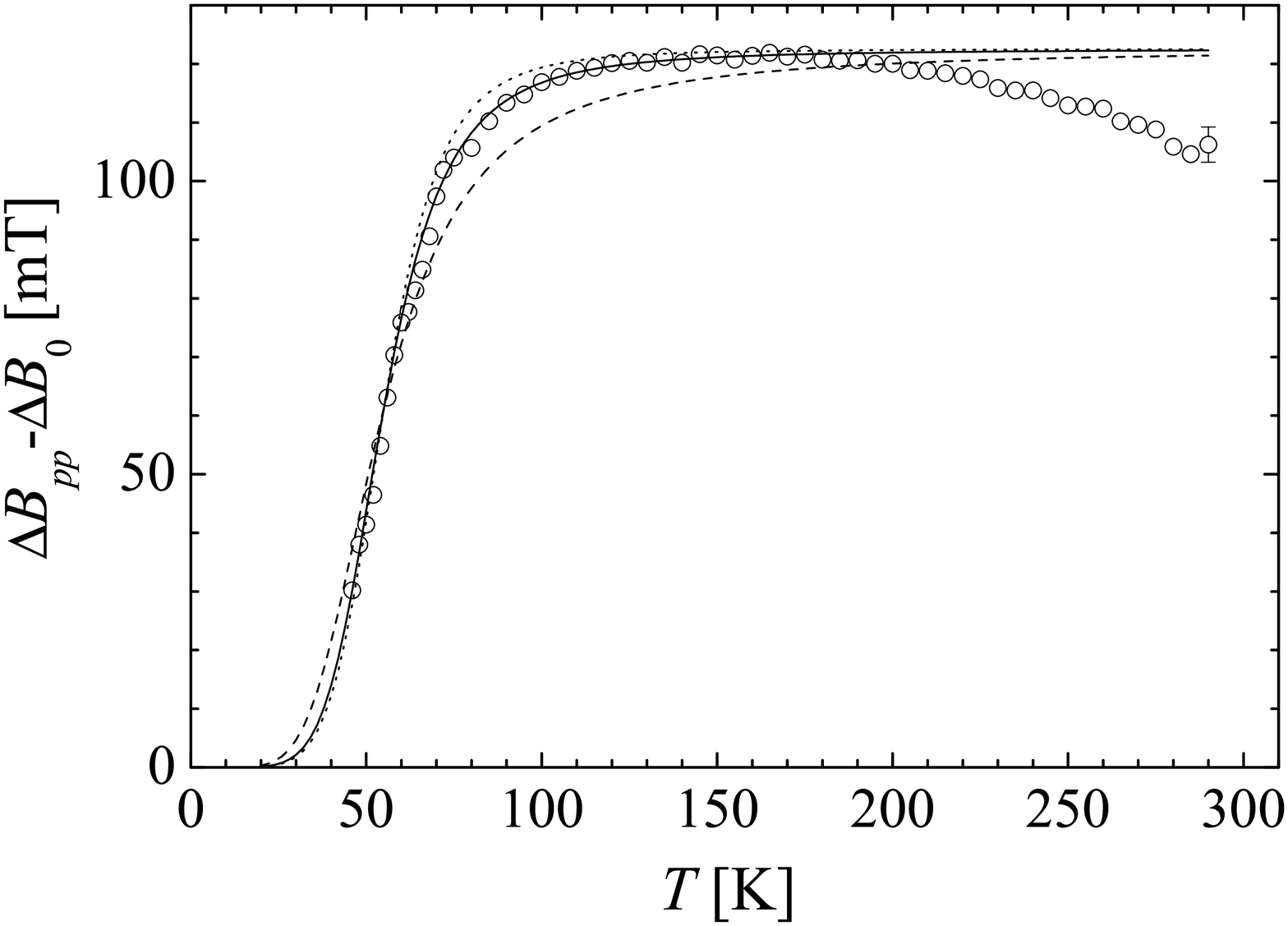}

\end{center}
\caption{ The bottleneck behavior of the impurity-induced linewidth
in PbNi${}_{1.98}$Co${}_{0.02}$V${}_{2}$O${}_{8}$. The curves represent
the temperature dependence predicted by the model Eqs. (\ref{XRef-Equation-43020556})-(\ref{XRef-Equation-42716040})
with the Debye temperature values $\theta _{D}=300$~K (dashed line),
$\theta _{D}=500$~K (solid line) and $\theta _{D}=700$~K (dotted
line).}
\label{XRef-Figure-427131011}
\end{figure}

\noindent where ${\Delta \textit{B}}_{0}$ represents the linewidth
of the pristine compound, ${\Delta \textit{B}}_{i}$ the impurity-induced
contribution to the linewidth and $\eta =\omega _{sl}/\omega _{e}^{i-h}$
the ratio between the rate of the impurity spin-lattice relaxation
and the impurity-host cross relaxation determined by the impurity-host
exchange {\itshape J}{\itshape ${}_{i-h}$}. A bottleneck effect
is expected to be observed since at low temperatures the spin-lattice
relaxation of the impurity spins is far below the impurity-host
cross relaxation.

As apparent from Fig. \ref{XRef-Figure-427131011}, the difference
between the linewidth of the pure PbNi${}_{2}$V${}_{2}$O${}_{8}$
and the PbNi${}_{1.98}$Co${}_{0.02}$V${}_{2}$O${}_{8}$ compounds
indeed increases with temperature from 45~K, which represents the
lowest temperature where the analysis is still reliable due to severe
broadening at lower temperatures (see Fig. \ref{XRef-Figure-427125441}).
The impurity-induced linewidth contribution reaches a broad maximum
of approximately ${\mathrm{\Delta }B}_{\mathit{i}}^{\mathit{max
}}=120$~mT around 150~K. The observed downturn at higher temperatures
will be discussed in the end of this subsection. Such dependence
is in line with the spin-lattice relaxation, which increases with
temperature due to the temperature-activated phonon density. Spin-lattice
relaxation of the impurities is expected to be due to Raman spin-phonon
processes and can be written as \cite{AbragamNMR}
\begin{equation}
{\left. \omega _{sl}=\frac{1}{T_{1}}=a\left( \frac{T}{\theta _{D}}\right.
\right) }^{7}\operatorname*{\int }\limits_{0}^{\theta _{D}/T}\frac{y^{6}
e ^{y}}{{\left(  e ^{y}-1\right) }^{2}}dy.%
\label{XRef-Equation-426163846}
\end{equation}

\noindent In Eq. (\ref{XRef-Equation-426163846}), parameter {\itshape
y} is defined as $y=\hbar \omega /k_{B}T$, $\theta _{D}$ represents
the Debye temperature and {\itshape a} is a constant depending on
the strength of the spin-phonon coupling. The latter two constants
are not known in our case, however, the {\itshape a} constant can
be conveniently dismissed from the expression for the expected linewidth
if we set $\eta ( 55~\mathrm{K}) =1$ as the experimental impurity
broadening reaches half of its saturated value at 55~K. In this
case the parameter $\eta ( T) $ can be expressed as
\begin{equation}
\eta ( T) =\frac{\omega _{sl}( T) }{\omega _{sl}( 55~\mathrm{K})
}.%
\label{XRef-Equation-42716040}
\end{equation}

The fit of the model described through Eqs. (\ref{XRef-Equation-43020556})-(\ref{XRef-Equation-42716040})
to the experimentally determined impurity-induced linewidth of the
ESR spectra is satisfactory for temperatures below 200~K. The best
agreement is reached when the Debye temperature is set to $\theta
=500$~K (solid line in Fig. \ref{XRef-Figure-427131011}). To see
the effect of this parameter on the quality of the fit, theoretical
curves with $\theta =300$~K (dashed line) and $\theta =700$~K (dotted
line) are also shown in Fig. \ref{XRef-Figure-427131011},
which allows us to determine the Debye temperature of the PbNi${}_{2}$V${}_{2}$O${}_{8}$
compound as $\theta =500(50)$~K.

Furthermore, the saturated value of the impurity-induced linewidth
contribution can be used for the estimation of the impurity-host
exchange interaction. When the isotropic exchange coupling is the
leading impurity-host interaction the
following equation can be derived in the high-temperature limit \cite{NagataJPSJ44}
\begin{equation}
\Delta  B_{i}^{max}=\frac{1}{\sqrt{3}}\frac{32}{{\textit{g}\mu }_{B}}\frac{J_{i-h}^{2}S_{i}(
S_{i}+1) }{3{\hbar \omega }_{e}}\frac{x}{2}.%
\label{XRef-Equation-427185133}
\end{equation}

\noindent Here {\itshape S}{\itshape ${}_{i}$} denotes the size
of the impurity spin, while the exchange frequency is defined as
${\hbar \omega }_{e}=\sqrt{M_{4}/M_{2}}$. In the case of the PbNi${}_{2}$V${}_{2}$O${}_{8}$
compound the single-ion anisotropy of the form $D\cdot S_{z}^{2}$
was reported to be the leading anisotropy term, with $D=-5.2$~K
determined from inelastic neutron scattering experiments \cite{ZheludovPRB62}.
The easy axis was found to point along the crystal {\itshape c}
axis (direction of spin chains). Using Eqs. (\ref{XRef-Equation-430231550}) and (\ref{XRef-Equation-43023168})
one can then derive the following expressions for the exchange frequency
$\omega _{e}=\sqrt{8}J/\hbar $, which predicts the impurity-host
exchange of the size $J_{i-h}=14$~K. The linear dependence of the
induced line broadening as predicted by Eq. (\ref{XRef-Equation-427185133})
is nicely revealed in Fig. \ref{XRef-Figure-4271361}b. The slope
of this line (6.8 T) makes an estimation of the impurity-host exchange
$J_{i-h}=15$~K, which is only slightly different from the above
value. Although the sign of $J_{i-h}$ can not be determined from
the above-presented ESR analysis, the magnetic susceptibility measurements
speak in favour of antiferromagnetic coupling, as argued in subsection
\ref{XRef-Section-52103829}.

The reduction of the Co-Ni exchange $J_{i-h}$ with respect to the
Ni-Ni intrachain exchange coupling {\itshape J} is expected. Namely,
it is well established that the strength of the antiferromagnetic
superexchange decreases with the reduction of the number of electrons
in the 3{\itshape d} orbital. For instance, the N\'eel temperature
of the NiO oxide with NaCl crystal structure reduces from 525~K
to 290~K in CoO oxide \cite{RadwanskiPhysicaB}. Another example
are the K${}_{2}${\itshape A}F${}_{4}$ compounds ({\itshape A} is
a transition metal ion), where the following exchange parameters
were obtained: $J_{\mathrm{Ni}}=102$~K, $J_{\mathrm{Co}}=16.8$~K,
$J_{\mathrm{Fe}}=15.7$~K, $J_{\mathrm{Mn}}=8.4$~K \cite{vanderVlistPRB30}.

At the end of this subsection, a short explanation of the downturn
of the impurity-induced linewidth above 150~K (see Fig. \ref{XRef-Figure-427131011})
is given. When the cobalt spin-lattice relaxation surpassed the
above evaluated exchange frequency of the pure system, the spin
fluctuations at nickel sites will become strongly affected by the
spin-lattice relaxation due to the strong exchange coupling with
cobalt ions \cite{GulleyPRB6}. For $\omega _{sl}\gg \omega _{e}$
the spin-lattice relaxation rate effectively substitutes the exchange
frequency in Eq. (\ref{XRef-Equation-427185133}) \cite{HoggSSC23},
which results in a narrowing of the absorption lines with raising
temperature due to the increase of the spin-lattice relaxation rate.
Such narrowing was experimentally observed in KMnF${}_{3}$ doped
with Fe${}^{2+}$ ions \cite{BialasBorgielJPC13}. If we approximate
the rate of the impurity-host cross relaxation as $\omega _{e}^{i-h}\approx
J_{i-h}/\hbar $ and take into account that $\omega _{sl}( 55~\mathrm{K})
=\omega _{e}^{i-h}$, the Eq. (\ref{XRef-Equation-426163846}) will
allow us to make an estimation that $\omega _{sl}>\omega _{e}$ for
$T>100$~K. This corresponds fairly well with the temperature of
150~K where ${\mathrm{\Delta }B}_{\mathit{i}}$ exhibits its maximum.
The effect is, however, not as drastic as one might naively expect,
which is due to the rather low concentration of Co${}^{2+}$ impurities.

\subsection{ESR measurements within the Haldane-gap regime}

At temperatures below the average Haldane gap $\Delta =43$~K 
of the PbNi${}_{2}$V${}_{2}$O${}_{8}$ system \cite{ZheludovPRB62},
in the Mg-doped compounds the tendency of line broadening with lowering
temperature suddenly alters (see Fig. \ref{XRef-Figure-427122636}).
In the mid-temperatures crossover regime, i.e., at $T\leqslant \Delta
$, the single almost Lorentzian resonance line speaks in favour
of a strong coupling of the liberated end-chain spins to the triplet
Haldane excitations. Similarly to the NMR results on Mg-doped compounds
\cite{ArconEPL65} the one-dimensional Haldane excitations are thus
again shown to coexist with the spin excitations emerging from the
end-chain spins.

The low-temperature signal, i.e., below 10~K where Haldane excitations
are severely suppressed due to the Haldane gap, can be exploited
to obtain an insight into the nature of the magnetic interactions
between the liberated spin degrees of freedom, which are responsible
for the occurrence of the magnetic ordering. As already emphasized
and again supported by the analysis presented in the subsequent
subsection, the coupling of the two spins neighboring a particular
vacancy site is ferromagnetic. Such ferromagnetic coupling effectively
produced an anisotropy Hamiltonian of the single-ion form $D^{*}{\tilde{S}}_{z}^{2}$,
where ${\tilde{S}}_{z}$ represents an effective spin operator of
the two-spin system. The size of the anisotropy in the case of the
uniaxial symmetry is given by \cite{ManakaPRB62}
\begin{equation}
D^{*}=-\frac{3\mu _{0}}{4\pi }\frac{{\left( g \mu _{0}\right) }^{2}}{r^{3}}-\frac{3}{2}{\left(
\frac{\Delta \textit{g}}{g}\right) }^{2}{\tilde{J}}^{\prime }.%
\label{XRef-Equation-42895649}
\end{equation}

\noindent In Eq. (\ref{XRef-Equation-42895649}) the first term arises
from the dipolar coupling of the two spins and the second one represents
the symmetric anisotropic exchange. The parameter ${\tilde{J}}^{\prime
}$ stands for the effective ferromagnetic coupling between the two
spins, which in our case is a result of the competing antiferromagnetic
next-nearest neighbor exchange of the pure chain, $J^{\prime }=5$~K, 
and the ferromagnetic coupling mediated through neighboring chains
\cite{ZheludovPRB64}. In addition, the antisymmetric anisotropic
exchange of the Dzyaloshinsky-Moriya type should also be included,
\begin{equation}
d\approx \left| \left( \Delta \textit{g}/g\right) {\tilde{J}}^{\prime
}\right| .%
\label{XRef-Equation-428103930}
\end{equation}

\noindent From the known lattice parameters \cite{MastorakiJSSC177}
the dipolar field at Ni${}^{2+}$ sites can be estimated, 
$B_{dd}=45$~mT. This value is approximately twice larger than maximum
internal fields observed in zero-field $\mu${}\textsuperscript{+}-SR
experiments \cite{LappasPRB66}. The estimated value of the dipolar
field is, however, far below the experimentally observed linewidths
at low temperatures. This fact favours the DM interaction as the
dominant broadening mechanism (first order in $\Delta\textit{g}/g$).

A plot of the linewidth of the ESR spectra at 5~K and 10~K reveals
a clear linear dependence upon the doping level (see
inset of Fig. \ref{XRef-Figure-427122636}). Such linear dependence
has been observed before in CuGeO${}_{3}$ and attributed to the
interacting delocalized spin clusters induced next to the impurities
\cite{GlazkovJETP93}. In the case of PbNi${}_{2}$V${}_{2}$O${}_{8}$
the spin clusters exhibit an exponential decay of spin polarization
with the correlation length $\xi \approx 6$ at $T=0$ \cite{KimEPJB4}.
Apart from the interaction of the two clusters neighboring a particular
spin vacancy and forming an effective spin $\tilde{S}=1$, there
is thus an additional sizable anisotropic interaction between neighboring
$\tilde{S}=1$ effective spins on the chain, which increases with the level 
of doping as reflected in the increase of the linewidth.

The zero-doping values of the linewidth can be assigned to be due
to the intrinsic anisotropy of the uncoupled effective $\tilde{S}=1$
impurity-induced spins. These values are virtually the same at 5~K 
and 10~K, i.e., ${\Delta \textit{B}}_{pp}^{x=0}=180$~mT, which
is in favour of the above assumption. Namely, due to the gapped
nature of magnetic excitations of the mediating Haldane medium,
the effective $\tilde{S}=1$ spins can be regarded as isolated in
the limit $x\rightarrow 0$. The increase of the linewidth below
10~K, which can be either due to the critical enhancement of antiferromagnetic
correlations close to the phase transition temperature \cite{ZorkoPRB65}
or a signature of the temperature evolution of the correlation length
\cite{KimEPJB4}, consequently does not affect the zero-doping linewidth.
This value can then be used for an estimation of the effective ferromagnetic
coupling within the effective $\tilde{S}=1$ spins, which is according
to Eq. (\ref{XRef-Equation-428103930}) of the size ${\tilde{J}}^{\prime
}=-2$~K. In this estimation the room-temperature value of $\Delta
\textit{g}/g=0.1$, which is not affected by static spin correlations,
was taken into account, as well as the fact that the exchange narrowing 
mechanism is not active in diluted magnetic systems \cite{AbragamNMR}.

\subsection{Susceptibility of the impurity induced in-gap states}\label{XRef-Section-52103829}

Using the above results of the exchange coupling between the impurity-induced
spins ${\tilde{J}}^{\prime }$ and impurity-host coupling
$J_{i-h}$, the temperature-dependence of the impurity-induced magnetic
susceptibility as shown in Fig. \ref{XRef-Figure-42710958}, can
also be quantitatively explained with a simplified model Hamiltonian.
In was shown by S{\o}rensen et al. \cite{SorensenPRB51} that the
full exchange Hamiltonian $\mathcal{H}_{e}+\mathcal{H}_{e}^{i-h}$
of the doped Haldane system at low doping levels can be replaced
by an effective Hamiltonian $\tilde{\mathcal{H}}$ for describing
the low-energy excitations. Following this approach the effective
Hamiltonians can be in our case written in the following form in
the limit of low cross-impurity exchange coupling ${\tilde{J}}^{\prime
}$ and low impurity-host exchange $J_{i-h}$ 
\begin{align}
\label{XRef-Equation-43023443}%
{\tilde{\mathcal{H}}}_{\mathrm{Mg}}&=\alpha {\tilde{J}}^{\prime
}{\textbf{S}}_{l}\cdot {\textbf{S}}_{r},
\\%
\label{XRef-Equation-43023534}%
{\tilde{\mathcal{H}}}_{\mathrm{Co}}&=\alpha  J_{i-h}( {\textbf{S}}_{l}\cdot
{\textbf{S}}_{i}+{\textbf{S}}_{i}\cdot {\textbf{S}}_{r}) +\alpha
{\tilde{J}}^{\prime }{\textbf{S}}_{l}\cdot {\textbf{S}}_{r},
\end{align}

\noindent where $\alpha =1.064$ \cite{SorensenPRB51}. Operators
${\textbf{S}}_{l}$, ${\textbf{S}}_{r}$ represent the $S=1/2$ effective
spins induced at sites ``left'' and ``right'' of a particular impurity
site reflecting the nature of the valence-bond-solid ground state
\cite{AffleckPRL59}. In the case of ${\tilde{J}}^{\prime },{\textit{J}}_{i-h}<J$,
bound in-gap states with exponentially decaying correlations are
predicted \cite{SorensenPRB51}.

Combining the effective exchange Hamiltonians given by Eqs. (\ref{XRef-Equation-43023443}),
(\ref{XRef-Equation-43023534}) and the Zeeman Hamiltonian of the
spins in the external magnetic field of 100~mT, the low-temperature
dependence of the uniform static susceptibility can be forecasted.
In the inset of Fig. \ref{XRef-Figure-42710958} the calculated values
of the expectation value of the {\itshape S}{\itshape ${}_{\mathit{z}}$}
operator multiplied by the level of doping is shown. This quantity
is detected in the dc magnetization measurements, however it reveals
the same information as the static susceptibility $\chi _{0}\propto
(\langle S_{z}^{2}\rangle -{\langle S_{z}\rangle }^{2})/k_{B}T$
as long as the magnetization vs. magnetic field curves are linear.
In the calculation the above-obtained parameters ${\tilde{J}}^{\prime
}=-2 $~K and $J_{i-h}=14 $~K were taken into account.
The agreement with the experiment is rather good, especially if
we recall that the impurity-induced susceptibility is given by 
\begin{equation}
\chi _{i}=\frac{x N_{A}g \mu _{B}\left\langle  S_{z}\right\rangle
}{2B}=11~\mathrm{emu}/\mathrm{mol}\cdot \frac{x}{2}\left\langle
S_{z}\right\rangle  .
\end{equation}

Although the experiment nicely follows the theoretical predictions,
it should be noted that close to the phase-transition temperature
the static spin-correlations effect makes the simple model picture
unsuited. Nevertheless, the puzzle of the size of the low-temperature
spin susceptibility in Co- and Mg-doped PbNi${}_{2}$V${}_{2}$O${}_{8}$
samples can be unambiguously unraveled. In the case of the vacancy
doping, the ferromagnetic coupling between the liberated end-chain
spins is responsible for the upturn in the magnetic susceptibility.
Here, it should be stressed that the size of the coupling constant
does not have any impact on the temperature dependence of the susceptibility.
On the other hand, antiferromagnetic coupling in the case of the
Co-doped sample is responsible for the substantial suppression of
the susceptibility in comparison to the Mg-doped compound, as it
shifts the weight of the in-gap states towards higher energies.
An anisotropic exchange is expected in the case of the Co${}^{2+}$
ions, which further broadens the calculated peak in magnetic susceptibility.

At the end, a short comment on Cu-doping is given. 
It was shown experimentally that the susceptibility curves
of Cu-doped PbNi${}_{2}$V${}_{2}$O${}_{8}$ were virtually indistinguishable from curves
corresponding to Mg-doped compounds above the phase-transition temperature
when the doping level of Cu${}^{2+}$ ions was twice the doping level
of Mg${}^{2+}$ ions \cite{UchiyamaPRL83}. This finding is in line
with our results. Namely, the antiferromagnetic exchange of Cu${}^{2+}$
impurities to Ni${}^{2+}$ spins is expected to have a high value,
so that the low-energy excitations are solely given by the ferromagnetic
coupling of the impurity-liberated spins. The ratio of the susceptibilities
thus reflects the ratio of the effective spins ${\tilde{S}}_{\mathrm{Mg}}=1$
and ${\tilde{S}}_{\mathrm{Cu}}=1/2$.

\section{Conclusions}

In this paper an explanation of the low-temperature magnetic properties
in magnetically and nonmagnetically doped Haldane compound
PbNi${}_{2}$V${}_{2}$O${}_{8}$ has been presented. The results of
the dc magnetization and the electron spin resonance measurements
were efficiently associated to provide an insight into the problem
of the impurity-induced long-range magnetic ordering. The ESR approach
revealed the strength of the ferromagnetic coupling between spin
degrees of freedom liberated at both sites of the vacancies in the
case of nonmagnetic Mg-doping, as well as the importance of the
magnetic interactions between these delocalized spin clusters. On
the other hand, we were also able to evaluate the impurity-host
exchange in compounds doped with magnetic Co${}^{2+}$ impurities
through the bottleneck-type of the impurity-induced broadening of
the ESR spectra. This coupling proved essential for the appearance
of the in-gap magnetic excitations. It allowed a rather accurate
prediction of the suppression of the low-temperature magnetic susceptibility
in Co-doped compounds through a simplified effective exchange model.
Additionally, the ESR measurements revealed information about the thermal vibrations in the
investigated system through the determination of the Debye
temperature.

\acknowledgments We thank the General Secretariat for Science \& Technology (Greece)
and the former Ministry of Education, Science and Sport of the Republic
of Slovenia for the financial support through a Greece-Slovenia
``Joint Research \& Technology Program''.

\appendix

\end{document}